\documentclass[aps,prb,twocolumn,groupedaddress,floatfix,showpacs]{revtex4-2}

\usepackage{graphicx}
\usepackage{amsmath}
\usepackage{subcaption}
\usepackage[T1]{fontenc}
\usepackage{color}

\begin{document}


\title{Superlens with copper and copper oxide}


\author{David Ziemkiewicz}
\email{david.ziemkiewicz@utp.edu.pl}

\author{Karol Karpi\'{n}ski}
\author{Sylwia Zieli\'{n}ska-Raczy\'{n}ska}

 \affiliation{Institute of
Mathematics and Physics, Technical University of Bydgoszcz,
\\ Al. Prof. S. Kaliskiego 7, 85-789 Bydgoszcz, Poland}


\date{\today}

\begin{abstract} 
We investigate the imaging properties of copper-based superlens surrounded by copper oxide (Cu$_2$O). A subwavelength image resolution of the order  $\lambda/9$ is demonstrated theoretically and  verified in numerical simulations. It is shown that the existence of excitons in Cu$_2$O influence the static and dynamical optical properties of the lens. In particular, an improvement of image quality caused by absorption in the spectral region of excitonic resonances is investigated. The plasmon-exciton interaction in the system may pave the way to a tunable, highly nonlinear superlens designs.
\end{abstract}


\maketitle
\section{Introduction}
In recent years optical communication through metallic nanostructures has attracted significant attention.
Since 2000 when Pendry initiated the idea of superlens imaging \cite{Pendry}, it has became an important research topic in past years. 
 Noble metal nanostuctures based on gold and silver have allowed the light to be guides beyond the diffraction limit, which has opened new possibilities for further miniaturization and applications of optoelectronic devices.
 
In 2005 Lee \textit{et al} \cite{Lee05} presented experimental and theoretical studies of the optical superlens using a thin silver slab and demonstrated optical imaging with resolution well below the diffraction limit.
The investigations of different, cheaper material to construct efficient superlens systems basing on different metals have started; Zhao \emph{et al} \cite{Zhao16} have studied several superlens materials, including  Ni, Cr  and also Cu in the 400 nm wavelength range concluding that the dielectrics covering the superlens surfaces have to be characterized by similar permittivities, which influence the imaging quality. It has also  been noted that a large permittivity of the dielectric in the front of the superlens is in favour of the imaging effect. Mkhitaryan \emph{et al} have recently demonstrated that high Q factor plasmonic structures can be realized with copper \cite{Mkhitaryan21}, further confirming that Cu is a promising candidate for construction of efficient superlens.

The general theoretical model valid for both homogeneous and
evanescent waves,  transfer-matrix analyses of the near-field metallic superlens  and
the numerical finite-difference in time-domain (FDTD) simulations has been proposed by Tremblay \cite{Tremblay} and the detailed
role of near-field optics in a superlens image formation has been reviewed recently by Adams et al \cite{Adams}.

Moreover, the important aspect of producing  superlens is the accomplishment of the optimal resolution. It is affected by the finite thickness of the superlens, which generates both long- and short-range surface plasmon modes, they in turn influence a transfer function distorting the image field. Wang et al \cite{Wang11} found that the presence of loss is not always detrimental to the superlens resolution. 
 
Since the discovery of negative index metamaterials and superlens, there is a strong need for a lens medium that is characterized not only by the necessary negative permittivity, but also one that offers tunability of its optical properties. By coupling various quantum systems with the metamaterial, one can provide such a capability \cite{Quach2011}. These so-called quantum metamaterials \cite{Stav2018} are a new class of media that bridges the gap between quantum systems and classical metamaterials. In this paper, we focus on the recently discovered Rydberg excitons (REs) in Cu$_2$O \cite{Kazimierczuk} and their potential application in copper based superlens.
REs are highly excited excitonic states characterized by a principal quantum number $n>>$1 for dipole-allowded P-type enveloped wavefunctions. In principle, they  have similar scaling as Rydberg atoms \cite{Heck17} although the physical origin of these similarities in their case bases on a complex valence band structure and different selection rules for excitons. The dimensions of REs reach micrometers and their life-times can reach hundreds of nanoseconds. Nowadays REs become one of the rapidly developing topic in solid-state optics; the studies extend from linear to nonlinear regimes in bulk media and nanostructures \cite{Orfanakis,my2022,Thomas2022}. Recently, REs interaction with plasmons has been studied, showing possibility of an evident extension of propagation length and bridging plasmonics and Rydberg excitons physics \cite{my_arxiv}. Here, we expand upon that work by considering the effect of excitons on the dynamics of copper-based superlens.

 We perform the detailed analysis of Cu and Cu$_2$O, with its particularly high $\epsilon=7.5$ based superlens taking into account all above mentioned aspects. 
 The required matching of permittivity of Cu and Cu$_2$O can be achieved either by tuning the frequency, adjusting the Cu film thickness \cite{Stenzel19} or using metal-dielectric composites \cite{Cai05} such as a fine mix of Cu and Cu$_2$O layers with excitons. Tailoring the optical properties of medium opened a new possibility of controlling the image. This is because the propagation dynamics of the field  depends on the diffraction and dispersion  properties of the medium. In particular, Archambault \emph{el al}\cite{Archambault12} pointed out that an improvement of spatial resolution has to be achieved using time-dependent incident pulse. In the spirit of their conclusions we examine the influence of the group velocities of various modes on the image formation. 
 
The paper is organized as follows. In the first section, a numerical model of the optical properties of copper and copper oxide is presented, which is later used in theoretical calculations and numerical simulations. Then, the system setup is outlined. The third section is devoted to the study of the steady state transmission coefficient of the copper superlens. Next, the dynamic of the image formation in the superlens is investigated. Finally, the conclusions are presented. The detailed description of the numerical simulation method is included in the appendix.

\section{Material model}
\begin{figure}[ht!]
\includegraphics[width=.9\linewidth]{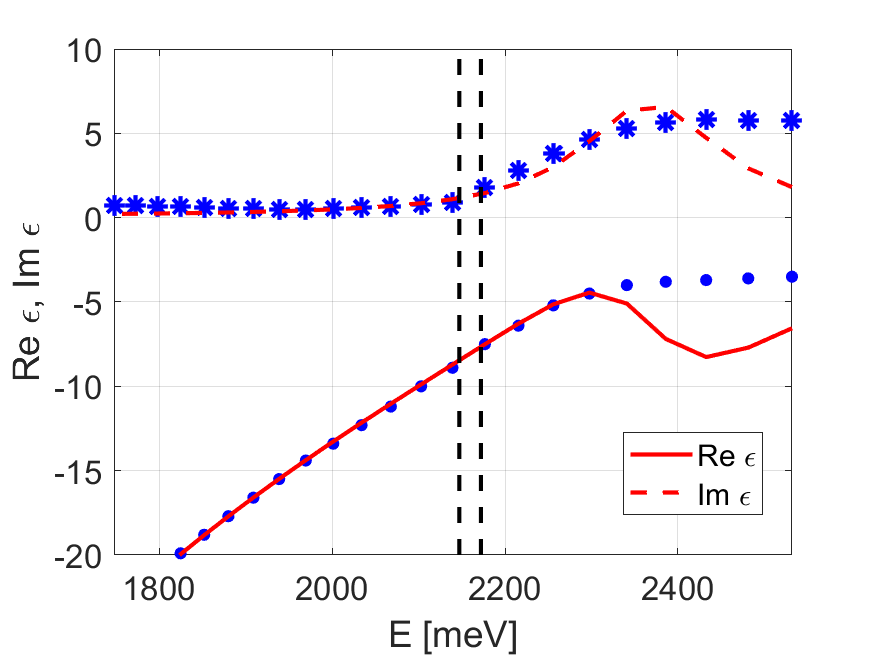}
\caption{Real (continuous line) and imaginary (dashed line) part of permittivity given by Eq. \ref{eq:DL}, compared to experimental data (dots).}\label{Cu_model}
\end{figure}

For the numerical description of the optical properties of Cu, we use the Drude-Lorentz model in which the permittivity, in the case of two-oscillator model, is given by \cite{Okada}
\begin{equation}\label{eq:DL}
\epsilon(\omega)=\varepsilon_\infty+\frac{f_0}{\omega_0^2-\omega^2+i\omega \Gamma_0}+\frac{f_1}{\omega_1^2-\omega^2+i\omega \Gamma_1}, 
\end{equation}
with the numerical value of parameters summarized in the Table \ref{parTable} in Appendix A. The values of parameters are adjusted to provide a good fit to the experimental data presented in ref. \cite{Hollstein}. A comparison of the experimentally measured permittivity spectrum and the model values is shown on the Fig. \ref{Cu_model}. In particular, the model is adjusted to provide a good fit in the energy region  between $E \sim 2140 - 2170$ meV (dashed, vertical lines), where excitonic resonances are present. In particular, we assume that the optical properties of Cu are determined by an ensemble of free electrons. The oscillator strength $f_0$ is determined by density of electron plasma, with so-called plasma frequency $\omega_p = \sqrt{f_0}$. Due to the fact that the electrons are not bound, there is no restoring force that would result in some intrinsic resonant frequency of charges and so $\omega_0=0$. The damping parameter $\Gamma_0$ represents various dissipative processes. The parameters of the second oscillator $f_1$, $\omega_1$, $\Gamma_1$ are used to include the effect of inter-band transitions that are responsible for an increase of absorption around 2.35 eV (see Fig. \ref{Cu_model}).

\begin{figure}[ht!]
\includegraphics[width=\linewidth]{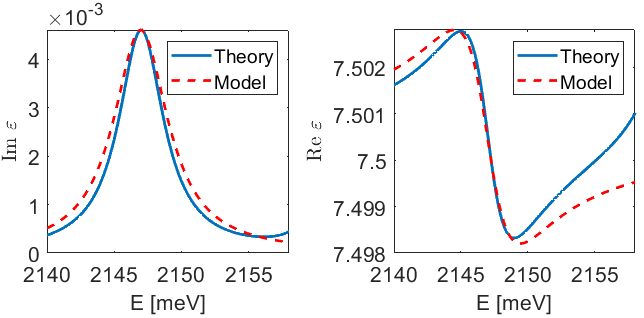}
\caption{Imaginary (left side) and real (right side) part of permittivity. Solid lines is given by theoretical calculation for Cu$_2$O bulk 2P exciton from \cite{Raczynska2016} and dashed lines given by Eq. \ref{eq:DL} with parameters depict in Table \ref{parTable}.}\label{Cu2O_model}
\end{figure}

For the copper oxide, the same Eq. (\ref{eq:DL}) with a different set of parameters can be used. The two main contributions to the Cu$_2$O susceptibility are the constant value of $\epsilon_\infty=7.5$ and the effect of 2P exciton resonance, as shown on the Fig. \ref{Cu2O_model}. This resonance is included as an oscillator with the frequency corresponding to the 2P exciton energy $2147$ meV and the oscillator strength and damping constant adjusted to match the height and width of absorption peak of the exciton \cite{Raczynska2016}. In the case of Cu$_2$O, the second oscillator in Eq. (\ref{eq:DL}) is unused.

\section{System setup}
\begin{figure}[ht!]
\includegraphics[width=0.6\linewidth]{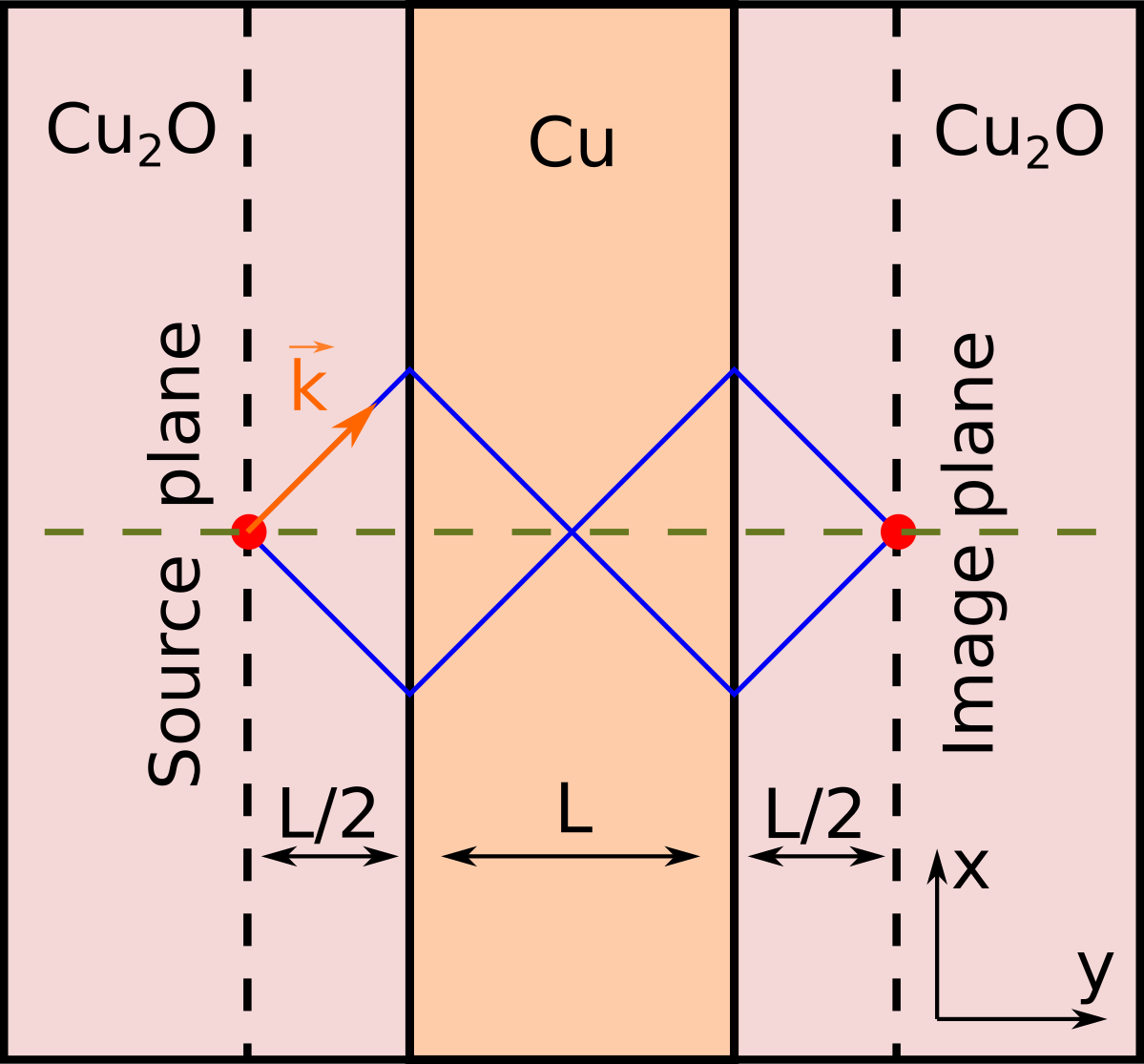}
\caption{Schematic representation of a copper superlens surrounded by Cu$_2$O.}\label{superlensmodel}
\end{figure}

Lets consider a multilayer system consisting of a thin copper layer surrounded by copper oxide, as shown on the Fig. \ref{superlensmodel}. The optical axis of the system (horizontal, dashed line) is aligned to the $y$ axis and the source/image planes (vertical dashed lines) are parallel to the $x$ axis. Blue lines depict a schematic of optical rays connecting the point source with its image (red dots). For an efficient superlens operation, the thickness $L$ needs to be smaller than approximately $0.1 \lambda$, where $\lambda$ is the wavelength of the light illuminating the system.  
The important parameter determining the superlens resolution is the wave vector $k_0$ and its $x$ component $k_x$. Specifically, to achieve subwavelength resolution, the superlens needs to transmit high spatial frequency components with $k_x>k_0$. In such a case, the $y$ component of the wave vector $k_y = \sqrt{k_0^2 - k_x^2}$ is imaginary and thus it describes evanescent waves. The key characteristic of the superlens is the capability to amplify these rapidly decaying evanescent waves \cite{Pendry}.

\section{Transmission coefficient}
One of the primary tools for studying the performance of the superlens is the transmission spectrum in the wave vector domain, e.g. the relation $T(k_x)$. The system is capable of subwavelength imaging when the values of $T(k_x>k_0)$ are significant. Similar to the approach presented in \cite{Wang11}, we calculate the transmission coefficient (transfer function) of a flat superlens in electrostatic limit, e.g. in the case where the lens thickness is considerably smaller than illuminated wavelength. Let's consider a source and image plane located at a distance $L/2$ from the superlens surfaces (see Fig. \ref{superlensmodel}). Transmission coefficient is calculated by multilayer transmission using Parratt recursion relation \cite{Parratt1954} 
\begin{equation}\label{eq:transmisja}
T=T_0T_1...T_N,
\end{equation}
where $T_j$ is transmission coefficient in each layer is given by 
 \begin{equation}
 T_j=e^{i k_z^{(j+1)}L_{j+1}} \frac{1+F_j}{1+F_jR_{j+1}},
\end{equation}
with reflection coefficient 
\begin{equation}
R_j=e^{2ik_z^{(j)}L_j}\frac{F_j+R_{j+1}}{1+F_j R_{j+1}},
\end{equation}
and 
\begin{equation}
F_j=\frac{\epsilon_{j+1}k_z^{j}-\epsilon_{j}k_z^{j+1}}{\epsilon_{j+1}k_z^{j}+\epsilon_{j}k_z^{j+1}}.
\end{equation}
Parameter $j=0,1,2$ numerates Cu$_2$O, Cu, Cu$_2$O layers, correspondingly. $k_z^{j}$ is z-component of wavevector in $j$ layer and $L_j$ is its thickness with $L_0=L_{N+1}=0$ and $R_{N+1}=0$. Similar approach one can found in other papers about superlenses, see \cite{Splawinski2021,Hakkarainen2009}

\begin{figure}[ht!]
\includegraphics[width=.9\linewidth]{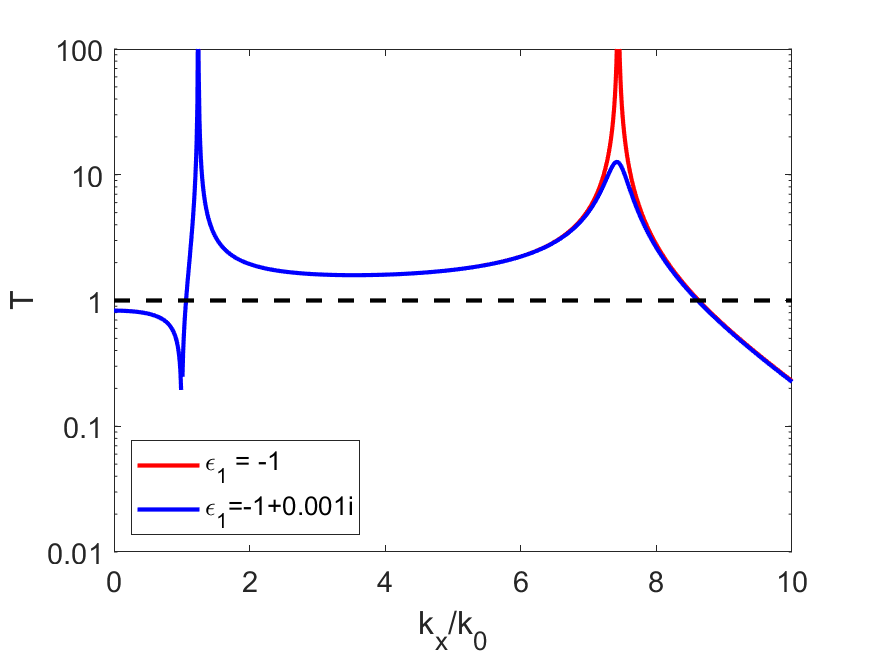}
\caption{Transmission coefficient as a function of the x component of wave vector $k_x$, calculated for a model superlens with $\epsilon=-1$.}\label{RT_verif}
\end{figure}

As a first step, we can verify the derived theoretical transmission relation by comparing it with the results presented by Wang \textit{at al} \cite{Wang11}. In particular, the Fig. \ref{RT_verif} depicts the transmission coefficient of idealized superlens consisting of a single layer of $\epsilon=-1$ material, with $\lambda/10$ thickness, surrounded by vacuum. There are several characteristic points in the transmission spectrum; for $k_x/k_0<1$, the transmitted mode corresponds to regular, propagating wave; the transmission coefficient in this range is smaller than 1. For higher values of $k_x$, the transmission coefficient describes evanescent waves that do not carry energy, so one can have $T>1$. In particular, the transmission spectrum usually contains two poles \cite{Wang11}, which are clearly visible on the Fig. \ref{RT_verif}. For $k_x/k_0>9$, there is a cutoff point where the transmission quickly drops. This means that the spatial resolution of the superlens is limited to approximately $\lambda/9$; one can increase this resolution by making the lens thinner \cite{Pendry}. The presence of the sharp poles in the transfer function (transmission coefficient) is generally undesirable as it introduces distortion \cite{Wang11}; one can see that the introduction of absorption (Fig. \ref{RT_verif}, blue line) significantly reduces the gain on the second, high $k_x$ pole, so that the transfer function becomes more flat (e.g. all wave modes are amplified to a similar degree). Thus, despite  the fact that absorption is in general detrimental to the superlens resolution, its presence can be beneficial to the image quality. Due to this reason, our proposal of Cu-based superlens has some advantages over traditional designs based on noble metals.

\begin{figure}[ht!]
\includegraphics[width=.9\linewidth]{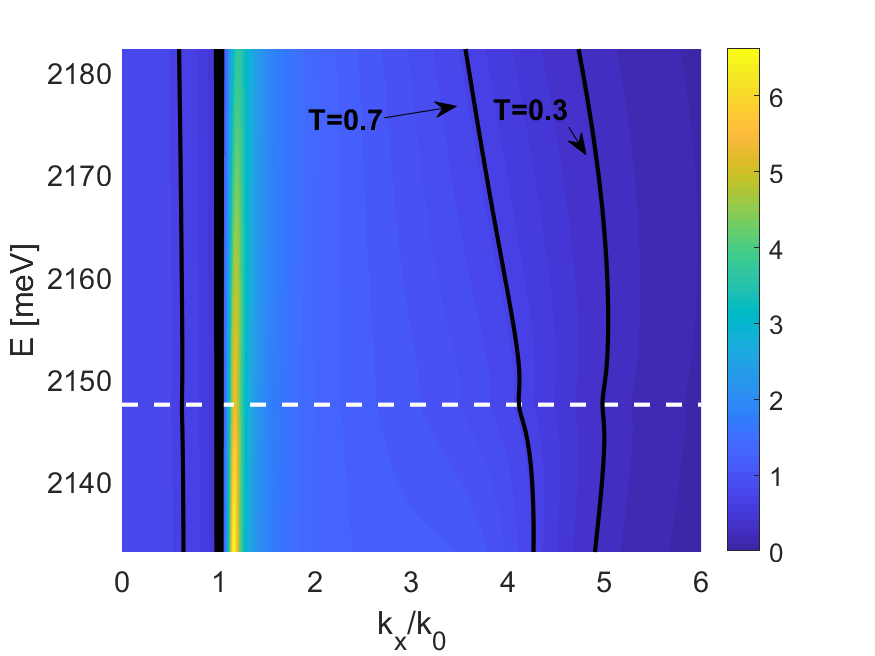}
\caption{Transmission coefficient (color) as a function of the x component of wave vector $k_x$ and energy.}\label{RT_1}
\end{figure}

As a next step, we consider a copper superlens with a thickness $L=20$ nm, surrounded by Cu$_2$O. From the perspective of practical applications, such a system is potentially simple to fabricate by controlled oxidation of Cu surface \cite{Rodriguez11}, vapour deposition \cite{Steinhauer20} or other techniques \cite{Lynch21}. The transmission coefficient calculated for a range of values of wave vector component $k_x$ and energy $E$ is shown on the Fig. \ref{RT_1}. 
Due to the fact that copper has a significantly higher absorption coefficient than the idealized superlens material considered previously, there is only a single pole in the transmission coefficient at $k_x \approx k_0$. Overall, the usable range of wave vector values where transmission $T > 0.1$ is $k_x < 6$, which corresponds to roughly $\lambda/7$ resolution and is comparable to silver superlens performance \cite{Sheng2009,Hakkainen2010}. One can notice that the strongest 2P exciton, which corresponds to 2147 meV (white, dashed line), produces a noticeable change in transmission coefficient (especially visible on the black contours). This means that the excitons in Cu$_2$O are a viable tool for fine-tuning of the lens properties in a very narrow energy range. 

\begin{figure}[ht!]
\centering
a)\includegraphics[width=.6\linewidth]{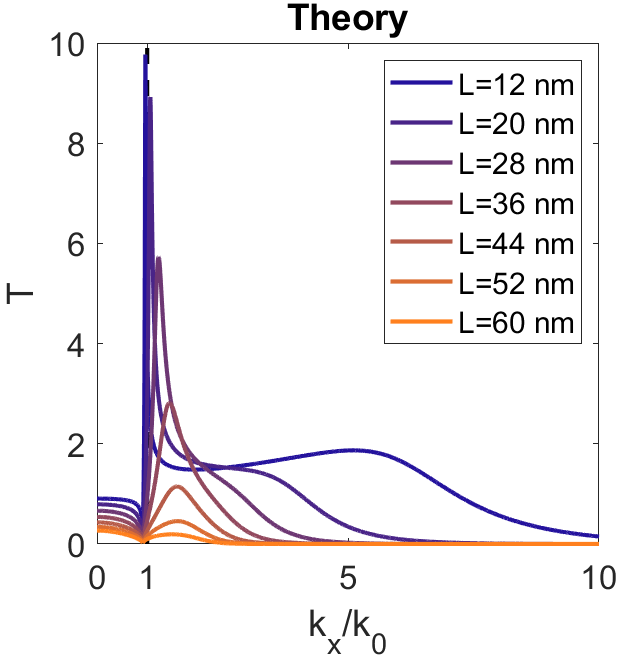}
b)\includegraphics[width=.6\linewidth]{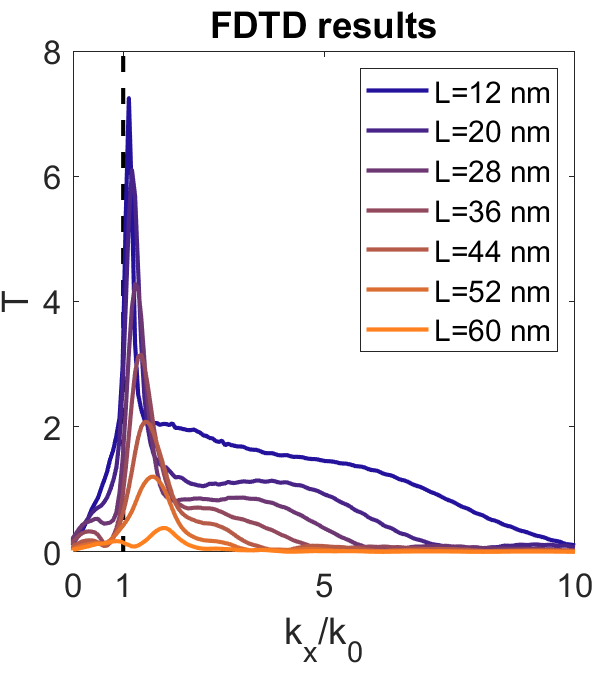}
\caption{Transmission coefficient as a function of the x component of wave vector $k_x$ a) calculated from Eq. (\ref{eq:transmisja}) and b) obtained in FDTD simulation.}\label{RT_2}
\end{figure}

One of the primary considerations in the superlens design is the choice of thickness. In fact in the case of metallic lens with negative permittivity and positive permeability, the thickness needs to be much smaller than illuminated wavelength \cite{Fang2003}. Also, in contrast to the ideal n=-1 superlens, there is a finite range of the values of wave vector component $k_x$ that are transmitted, as illustrated earlier on the Fig. \ref{RT_1}. In general, we can expect that high spatial-frequency components of the image decay much faster in a thicker lens \cite{Wang11}. This is indeed the case in the calculated results shown on the Fig. \ref{RT_2}. Overall, there is a good match between theoretical and numerical results. The first local transmission maximum, which is also present in the idealized case on the Fig. \ref{RT_verif}, forms a very sharp peak in theoretical calculations. To further confirm these results, we use the Finite-Difference Time-Domain (FDTD) simulation that is based directly on Maxwell's equations and thus allows for a  direct study of superlens optical properties with no simplifying assumptions that are present in theoretical description. The only important limitation of FDTD is the limited time and spatial resolution, which results in a smaller, broader peak in the numerical transmission spectrum (Fig. \ref{RT_2} b)). As mentioned before, the secondary peak on the Fig. \ref{RT_verif} located around $k_x=7k_0$ is very sensitive to absorption and in case of Cu, it is almost completely suppressed. One can see that only in the case of $L=12$ nm, a very wide local maximum of transmission is visible on the Fig. \ref{RT_2} at $k_x \approx 5 k_0$. Similar to the earlier results, the lack of sharp peaks in the transmission spectrum is beneficial, as it means that all spatial frequency components are amplified to the same degree \cite{Wang11}. In general, a lens thickness $L$ of the order of 10 nm is favourable, with significant transmission up to $k_x/k_0 \approx 10$. As the lens becomes thicker, the transmission cutoff becomes smaller, up to $k_x/k_0 \approx 2$ for the thickest considered lens $L=60$ nm.

After establishing the accuracy of the FDTD simulation, one can use the numerical results to investigate various aspects of superlens operation that are not described by the steady state transmission coefficient. As mentioned in the discussion of the Fig. \ref{RT_1}, the presence of excitons has a limited impact on transmission coefficient; overall, the imaginary part of Cu$_2$O permittivity in the spectral range of 2P exciton resonance is of the order of $Im$ $\epsilon \sim 10^{-3}$ (Fig. \ref{Cu2O_model}) as compared to $Im$ $\epsilon \sim 1$ in the case of copper (\ref{Cu_model}). Therefore, the majority of absorption comes from the metal. On the other hand, it is well known that the surface plasmons-polaritons (SPPs) excited on the metal-dielectric interface are very sensitive to the susceptibility changes in the dielectric. In the case of superlens, these plasmons are excited on both metal-dielectric interfaces and they play a pivotal role in the evanescent field amplification and thus the image formation \cite{Pendry,Smith03}. At the same time, they tend to distort the image field \cite{Wang11}. 

\begin{figure}[ht!]
\centering
a)\includegraphics[width=.9\linewidth]{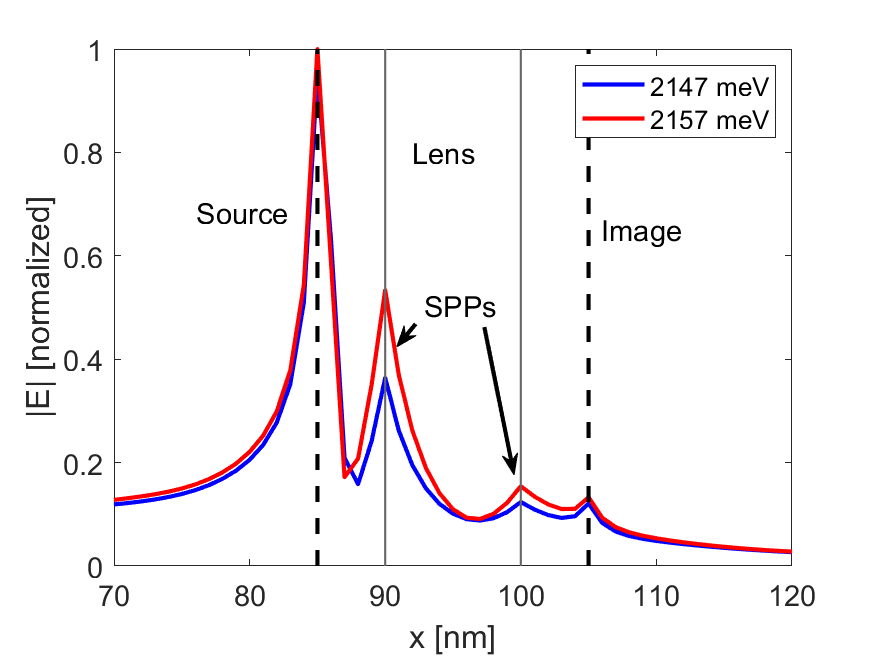}
b)\includegraphics[width=.9\linewidth]{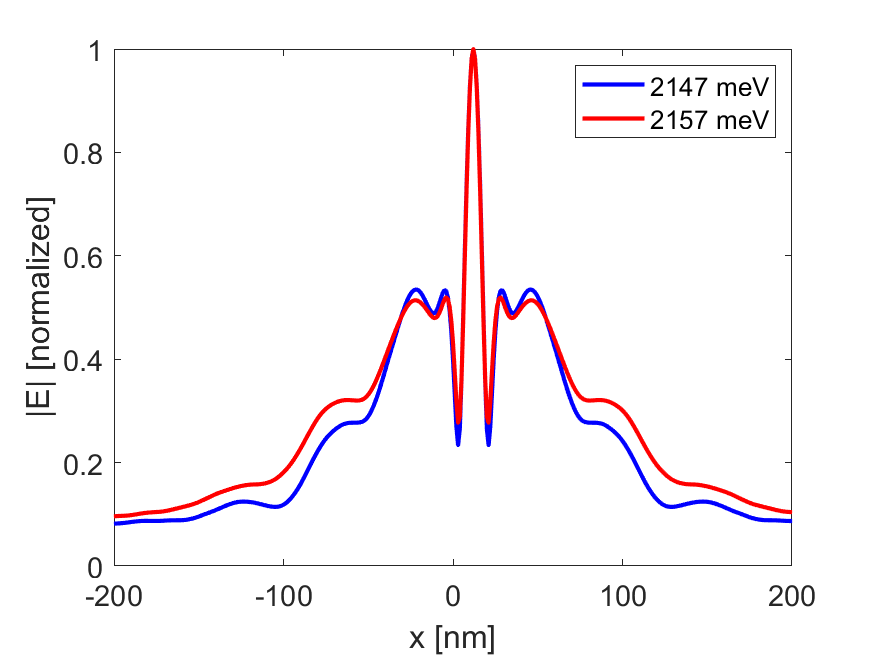}
\caption{Normalized electric field amplitude a) in the optical axis b) in the image plane.}\label{RT_3}
\end{figure}

To study these plasmons and their influence on superlens operation, let's consider the numerically obtained electric field amplitude in the optical axis of the system e.g., on the line connecting a single point source and its image, shown on the Fig. \ref{RT_3} a). Naturally, the highest peak of the field amplitude is located at the source. One can also see significant peaks on the Cu-Cu$_2$O interfaces (continuous, vertical lines). These maxima correspond to surface plasmon resonances. The last peak is located on the image plane. Interestingly, while the image peak is almost the same regardless whether the energy is tuned to the 2P excitonic resonance (2147 meV) or not, the local maxima corresponding to SPPs are significantly affected by the presence of excitons. This means that the superlens containing excitons is capable of reconstructing the image while partially suppressing the surface plasmons. As a result, in the image of the point source (Fig. \ref{RT_3} b)), the side maxima caused by interference with the plasmons are reduced. These findings are consistent with the results by Wang \textit{et al} \cite{Wang11}, where it is shown that adding loss in the image region can provide a significant enhancement of the image quality.
It should be noted that the influence of excitons on the system is limited to a very narrow part of the spectrum; the largest 2P exciton resonance is characterized by a linewidth of $\Gamma \sim 5$ meV, while the energy range where a good match between Cu$_2$O and copper permittivity that allows for superlens operation is wider than the whole excitonic spectrum (see Fig. \ref{Cu_model}). This means that the system can be fine tuned with small changes of illuminating light frequency without impacting the image quality due to mismatch of permittivities.   

\section{Superlens in time domain}

\begin{figure}[ht!]
\includegraphics[width=.999\linewidth]{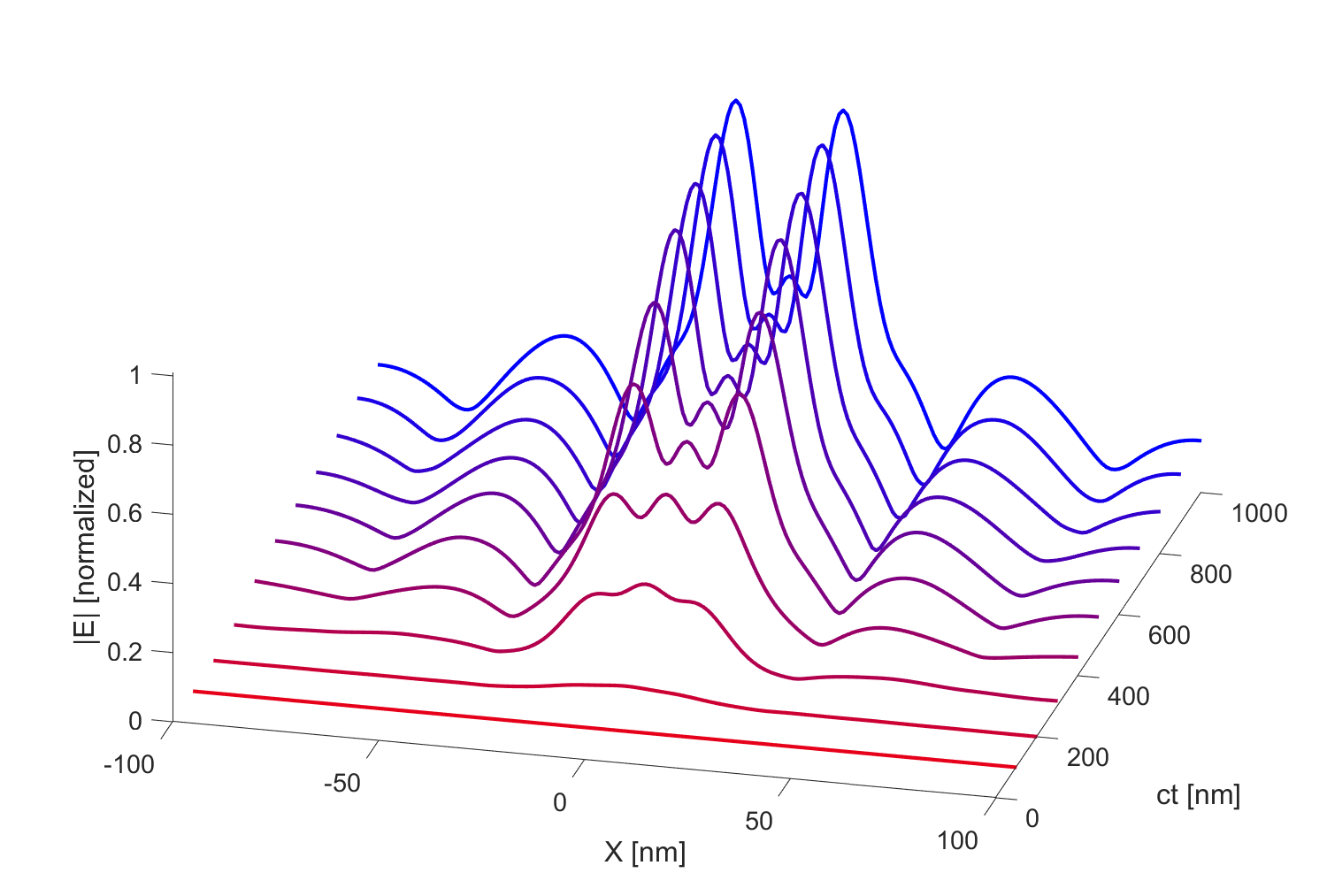}
\caption{Normalized field amplitude in the image plane as a function of time.}\label{EVO1}
\end{figure}

An important aspect of the comprehensive description of the superlens operation is the study of the time evolution of the image. The Fig. \ref{EVO1} depicts the field amplitude in the image plane of a $L=20$ nm superlens illuminated by two point sources. The distance between sources is 40 nm, which is approximately $\lambda/5$ (the energy of 2147 meV corresponds to a wavelength of 571 nm in vacuum and 210 nm in Cu$_2$O). The time unit is scaled by the speed of light in Cu$_2$O  which is around $ 1.1 \cdot 10^8$ m/s. 
One can see that initial, wide peak forming in the center of the image quickly transforms into two narrow peaks that are images of the radiation sources. The gradual formation of the image is caused by the varying group velocity of specific wave modes. In particular, the high $k_x$ modes that correspond to the fine details of the image are characterized by a low group velocity and thus need longer time to propagate through the lens. Alternatively, one can describe this process in terms of mode amplification; an evanescent wave with very large $k_x$ (e.g large, imaginary wave vector) decays very rapidly in free space. Thus, to reconstruct the image, these modes need to be more amplified by the lens, which takes longer time. 

\begin{figure}[ht!]
\includegraphics[width=.9\linewidth]{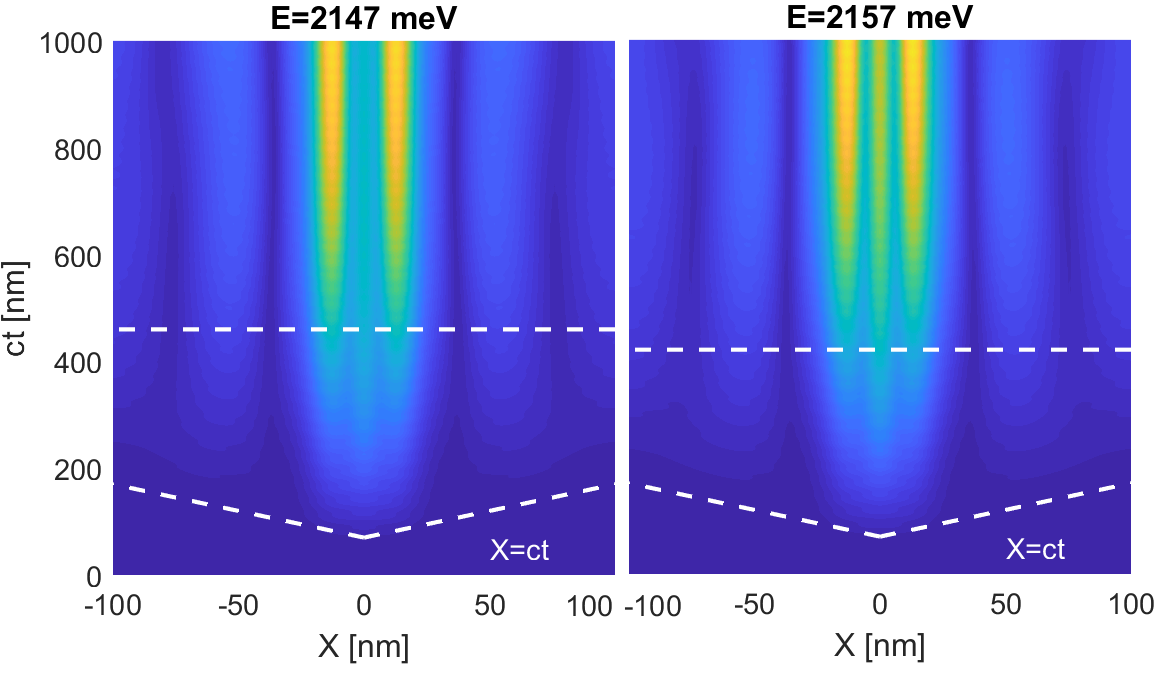}
\caption{Normalized field amplitude in the image plane (color) as a function of time, for two different energies.}\label{EVO2}
\end{figure}

On the Fig. \ref{EVO2}, a similar evolution is presented, calculated for two energies. On the left panel the energy is set to match the 2P exciton energy, while the right panel corresponds to the spectral region away from excitonic resonances.
The diagonal dashed lines on the bottom correspond to the casualty limit; no field can propagate in the system faster than $x=ct$. The horizontal dashed line corresponds to the moment when the image field reaches half of its maximum amplitude. In the case where excitonic resonance is present, the group velocity of wave modes is reduced due to the fact that an excitonic resonance corresponds to a sudden increase of normal dispersion. Specifically, we recall that the group velocity is given by
\begin{equation}
V_g = \frac{c}{n+\omega\frac{\partial n}{\partial \omega}}
\end{equation}
where $\omega$ is the radiation frequency and $n=\sqrt{\epsilon}$ is the refraction coefficient. In the spectral range of excitonic resonance, the permittivity changes very rapidly with frequency, so that $\partial n/\partial \omega$ has a significant value and noticeably reduces the group velocity \cite{my_arxiv}. Due to this, the image formation on the left panel of the Fig. \ref{EVO2} is slightly delayed. Notably, the presence of excitons has a stabilizing effect on the image; the spurious peak located between the two source images is noticeably weaker in the left panel, confirming the earlier findings that additional absorption in dielectric is beneficial to the image quality.

It should be also noted that the superlens resolution can be potentially greatly improved by considering time-dependent illumination \cite{Archambault12}. The use of excitons opens up a new opportunity to provide time-dependent superlens properties; by using a separate control field, one can take advantage of the nonlinear properties of excitons in Cu$_2$O due to the Rydberg blockade effect and change its absorption on demand. Furthermore, we recall that the optical properties of excitons are highly nonlinear \cite{my2022}; in the case of superlens, the optical field is highly focused in subwavelength volume, further enhancing the system nonlinearity, especially in high power pulsed operation. Finally, one can use an external field to modulate the spatial distribution of excitons and thus induce a spatially nonuniform optical properties of Cu$_2$O. In such a case, it might be possible to realize enhanced transmission of evanescent waves and improvement of resolution that are usually accomplished with lens surface roughness \cite{Fang2003, Wang11}. In contrast with fabricated roughness, the exciton distribution can be tuned on demand.  

\section{Conclusions}
Since their recent discovery in 2014, Rydberg excitons in Cu$_2$O are becoming a very valuable tool in creating nonlinear, controllable optical nanostructures. One of their possible applications are tunable plasmonic devices, where copper is used as the metal that facilitates excitation of SPPs and allows for easy fabrication of the structure. We show that a Cu layer surrounded by Cu$_2$O can act as an efficient superlens. It is demonstrated that such a device is capable of subwavelength imaging with a resolution approaching $\lambda/9$, which is competitive with traditional silver and gold based setups. Moreover, it is shown that excitons in Cu$_2$O can be used to modify the optical properties of the lens on demand and that the increased absorption due to the excitonic resonances is beneficial to the image quality. Several possible prospects for taking advantage of nonlinear optical properties of Cu$_2$O due to the excitons are outlined in the context of superlens design.

\section{Acknowledgments}
Support from the National Science Centre, Poland (NCN), project
Miniatura, 2022/06/X/ST3/01162, is greatly acknowledged.

\appendix
\section{FDTD method}
The performed simulations are based on a standard Yee algorithm \cite{Taflove}, where a set of field evolution equations is derived from Maxwell's equations and used to update the electric/magnetic field values within some defined volume (computation domain), with a fixed time step. The whole domain is divided by a rectangular grid with a single cell size $\Delta x$. A two-dimensional system with TM field configuration is chosen for simplification of calculation; the electric field has two components in the plane of the propagation $\vec{E}=[E_x,E_y,0]$, and the magnetic field has a single component perpendicular to the $xy$ plane $\vec{H}=[0,0,H_z]$. The two-dimensional representation is valid as long as the represented system (Fig. \ref{superlensmodel}) is much larger than the wavelength in the z direction. At every grid point, the electric and magnetic field distributions  are calculated from their previous values with evolution equations derived directly from Maxwell's equations. 

To model the dispersive properties of the optical media in the simulation, the auxiliary differential equation (ADE) approach is used \cite{Okada}. The medium is characterized by a polarization vector $\vec{P} = [P_x, P_y, 0]$. The evolution of polarization is described by a second-order partial differential equation in the form 
\begin{equation}
\ddot{P}+\Gamma_j \dot{P}+\omega_j=\frac{\epsilon_0 f_j}{{\epsilon_\infty}}\vec{E},
\end{equation}
where, depending on the medium complexity, several oscillatory terms $j=1,2,3...$ can be used with fitted parameters $f_j$, $\Gamma_j$, $\omega_j$ and $\epsilon_\infty$ is the constant (frequency independent) part of permittivity. The full set of equations solved in our FDTD approach is as follows   
\begin{eqnarray}
\frac{\partial H_z}{\partial t}&= \frac{1}{\mu_0}\left(\frac{\partial E_x}{\partial y}-\frac{\partial E_y}{\partial x} \right),\\
\frac{\partial E_x}{\partial t}&= \frac{1}{\epsilon_0}\left(\frac{\partial H_z}{\partial y}-\frac{\partial P_x}{\partial t} +j_x\right),\\
\frac{\partial E_y}{\partial t}&= \frac{1}{\epsilon_0}\left(\frac{\partial E_x}{\partial y}-\frac{\partial P_y}{\partial t}+j_y \right),
\end{eqnarray} 
where $\mu_0$, $\epsilon_0$ are the vacuum permittivity and permeability, respectively, and $j_x$,$j_y$ are components of source current density. The above equations are rearranged to obtain time derivatives of the $E_x$, $E_y$, $H_z$ fields, which are then used to calculate the field evolution with some constant time step $\Delta t$. Unit normalisation is used so that $\mu_0 = \epsilon_0 = c = 1$. The parameters of Drude-Lorentz model of copper and copper dioxide, which are used in simulation are shown in Table \ref{parTable}
\begin{table}[ht] 
\centering
\footnotesize
\begin{tabular}{|l|l|l|l|l|l|l|l|}
    \hline
        Medium & $\epsilon_\infty$ & $f_0$ [eV$^2$] & $\omega_0$ [eV] & $\Gamma_0$ [eV] & $f_1$  [eV$^2$] & $\omega_1$ [eV] & $\Gamma_1$ [eV] \\ \hline
        Cu$_2$O & 7.5 & 489$\cdot 10^{-7}$ & 2.147  & 0.0049 & 0 & 0 & 0 \\ \hline
        Cu & 15.3 & 122.18 & 0 & 0.0025 & 3.18 & 2.37 & 0.20 \\ \hline
    \end{tabular}
    \caption{Parameters of Cu fitted in Drude-Lorentz model to experimental \cite{Hollstein}. The Cu$_2$O fitted to bulk experimental data \cite{Kazimierczuk}, considering only 2P-exciton.}\label{parTable}
\end{table}  

The simulation space is divided into a 300 $\times$ 1024 grid points. Computation domain is surrounded by absorbing boundaries to reduce reflections. The spacial cell size $\Delta x =$ 4nm and time step $\Delta t=6.67 \cdot 10^{-18}$ s. The total simulation time is on the order of $10^{-14}$ s which is sufficient for the surface plasmons excited on the lens boundaries to reach steady state amplitude. At the same time, the optical properties of excitons are effectively static on this time scale (e.g. exciton density is constant, excitons are not moving, there is no exciton formation/recombination during the simulation). The lens system is placed in the middle of domain with optical axis in the x direction. A point source of radiation placed in source plane as a current $j_y$ with sinusoidal time dependence. In the calculation of reflectivity spectrum, the field amplitude distribution in the object and image planes (Fig. \ref{superlensmodel}) is calculated after steady state operation is established. Then, a Fourier transform is performed to obtain the field distribution as a function of the wavevector component $k_x$.   

Data in source and image planes has been collected from 2000 to 8000 time step number. Afterwards Fourier analysis was perform in the space domain  to obtain field distribution as a function of the wavevector component $k_x$, and as a result transmission coefficient. In the image evolution analysis, a set of spatial field distributions is recorded at various times.


\begin{thebibliography}{99}
\bibitem {Pendry}
J. B. Pendry, ,,Negative Refraction Makes a Perfect Lens'', Phys. Rev. Lett. \textbf{85}, 16, 3966 (2000).

\bibitem{Lee05}
H. Lee, Y. Xiong, N. Fang, W. Srituravanich, S. Durant, M. Ambati, C. Sun and X. Zhang, ,,Realization of optical superlens imaging below the diffraction limit'' New Journal of Physics \textbf{7}, 255 (2005). 

\bibitem{Zhao16}
C. Zhao, Y. Zhou, Y.Zhang, H. Wang, ,,The imaging properties of the metal superlens'', Optics Communications \textbf{368}, 180-184 (2016). 


\bibitem{Mkhitaryan21}
V. Mkhitaryan, K. March, E. Tseng, X. Li, L. Scarabelli, L. Liz-Marz\'{a}n, S. Chen, L. H. G. Tizei, O. St\'{e}phan, J. Song, M. Kociak, F. Garcia de Abajo, and A. Gloter, ,,Can Copper Nanostructures Sustain High-Quality Plasmons?'', Nano Lett. \textbf{21}, 2444-2452 (2021).


\bibitem{Tremblay}
G. Tremblay, Y. Shen, ,,Modeling and designing metallic superlens with metallic objects'',  Optics Express \textbf{19}, 21, 20634-20641 (2011). 

\bibitem{Adams}
W. Adams, M. Sadatol, and D.O. Guney, ,,Review of near-field optics and superlenses for sub-diffraction-limited nano-imaging'', API Advances \textbf{6}, 100701 (2016).

\bibitem{Wang11}
H. Wang, J. Bagley, L. Tsang, S. Huang, K. Ding, and A. Ishimaru, ,,Image enhancement for flat and rough film plasmon superlenses by adding loss'', J. Opt. Soc. Am. B \textbf{28}, 10, 2499-2509 (2011).

\bibitem{Quach2011}
J. Quach, C. Su, A. Martin, A. Greentree, and L.Hollenberg, ,,Reconfigurable quantum metamaterials'', Optics Express \textbf{19}, 12, 11018-11033 (2011).

\bibitem{Stav2018}
T. Stav, A. Faerman, E. Maguid, D. Oren, V. Kleiner, E. Hasman, M. Segev, ,,Quantum entanglement of the spin and orbital angular momentum of photons using metamaterials'', Science 361(6407), 1101-1104 (2018).

\bibitem {Kazimierczuk}
T. Kazimierczuk, D. Fr\"{o}hlich, S. Scheel, H. Stolz, and M. Bayer, ,,Giant Rydberg Excitons in Cuprous Oxide'', Nature \textbf{514}, 344 (2014).

\bibitem{Heck17}
J. Heck\"{o}tter, M. Freitag, D. Fröhlich, M. Assmann, M. Bayer, M. A. Semina, and M. M. Glazov, ,,Scaling laws of Rydberg excitons'', Phys. Rev. B \textbf{96}, 125142 (2017).

\bibitem{Orfanakis}
K. Orfanakis, S.K. Rajendran, H. Ohadi, S. Zielinska-Raczynska, G. Czajkowski, K. Karpinski, and D. Ziemkiewicz, ,,Quantum confined Rydberg excitons in Cu$_2$O nanoparticles''
Phys. Rev. B 103, 245426  (2021).

\bibitem{my2022}
D. Ziemkiewicz, Gerard Czajkowski, Karol Karpi\'{n}ski, and Sylwia Zieli\'{n}ska-Raczy\'{n}ska, ,,Nonlinear optical properties and Kerr nonlinearity of Rydberg excitons in Cu$_2$O quantum wells'',
Phys. Rev. B \textbf{106}, 085431, (2022).

\bibitem{Thomas2022}
C. Morin, J. Tignon, J. Mangeney, S. Dhillon, G. Czajkowski, K. Karpi\'{n}ski, S. Zieli\'{n}ska-Raczy\'{n}ska, and D. Ziemkiewicz, T. Boulier, ,,Self-Kerr Effect across the Yellow Rydberg Series of Excitons in Cu$_2$O'' Phys. Rev. Lett. \textbf{129}, 137401, (2022).

\bibitem{my_arxiv}
D. Ziemkiewicz, S. Zieli\'{n}ska-Raczy\'{n}ska, ,,Copper plasmonics with excitons'', arXiv:2208.07184. 

\bibitem{Stenzel19}
O. Stenzel, S. Wilbrandt, S. Stempfhuber, D. G\"{a}bler, and S. Wolleb, ,,Spectrophotometric Characterization of Thin Copper and Gold Films Prepared by Electron Beam Evaporation: Thickness Dependence of the Drude Damping Parameter'', Coatings \textbf{9}, 181 (2019).

\bibitem{Cai05}
W. Cai, D. Genov, and V. Shalaev, ,,Superlens based on metal-dielectric composites'', Phys. Rev. B \textbf{72}, 193101 (2005).

\bibitem{Archambault12}
A. Archambault, M. Besbes, and J. Greffet, ,,Superlens in the Time Domain'', PRL 109, 097405 (2012).

\bibitem{Okada}
N. Okada and J.B. Cole, ,,Effective Permittivity for FDTD Calculation of Plasmonic Materials'', Micromachines  \textbf{3}, 168-179 (2012).

\bibitem{Hollstein}
T. Hollstein, U. Kreibic, F. Lens, ,,Optical Properties of Cu and Ag'', Phys. Stat. Sol. (b) \textbf{83}, 545 (1977).

\bibitem{Raczynska2016}
S. Zieli\'{n}ska - Raczy\'{n}ska,  G. Czajkowski,D. Ziemkiewicz, ,,Optical properties of Rydberg excitons and polaritons'', Phys. Rev. B 93, 075206 (2016). 

\bibitem{Parratt1954}
L. G. Parratt, ,,Surface Studies of Solids by Total Reflection of X-Rays'', Phys. Rev. \textbf{95}, 2, 359 (1954).

\bibitem{Splawinski2021}
M. Splawinski, S. Bostock, K. J. Chau, and L. Markley, ,,Superlens coupling to object and image: A secondary resonance mechanism to improve single-negative imaging of electromagnetic waves'', J. Appl. Phys. 129, 163102 (2021). 

\bibitem{Hakkarainen2009}
T. Hakkarainen, T. Set\"{a}l\"{a}, A. T. Friberg, ,,Subwavelength electromagnetic near-field imaging of point dipole with metamaterial nanoslab'', J. Opt. Soc. Am. A 26, 2226 (2009).

\bibitem{Rodriguez11}
O. Peña-Rodríguez, U. Pal, ,,Effects of surface oxidation on the linear optical properties of Cu nanoparticles'', J. Opt. Soc. Am. B \textbf{28}(11), 2735-2739 (2011).

\bibitem{Steinhauer20}
S. Steinhauer, M. A. M. Versteegh, S. Gyger, A. W. Elshaari, B. Kunert, A. Mysyrowicz, and V. Zwiller, ,,Rydberg excitons in Cu2O microcrystals grown on a silicon platform'', Nature Communications Materials \textbf{1}, 11 (2020).

\bibitem{Lynch21}
S. A. Lynch, C. Hodges, S. Mandal, W. Langbein, R. P. Singh, L. A. P. Gallagher, J. D. Pritchett, D. Pizzey, J. P. Rogers, C. S. Adams, and M. P. A. Jones, ,,Rydberg excitons in synthetic cuprous oxide Cu$_2$O'', Phys. Rev. Materials \textbf{5}, 084602 (2021). 

\bibitem{Sheng2009}
G. Tremblay, Y. Sheng, ,,Imaging performances of the metallic superlens'', Proc. of SPIE Vol. 7156 715602-1 (2009).

\bibitem{Hakkainen2010}
T. Hakkarainen, T. Set\"{a}l\"{a}, A.T. Friberg, ,,Near-field imaging of point dipole with silver superlens'', Appl Phys B (2010) 101: 731-734 (2010).

\bibitem{Fang2003}
N. Fang, Z. Liu, T. Yen, and X. Zhang, ,,Regenerating evanescent waves from a silver superlens'', Optics Express \textbf{11}, 7, 682-687 (2003). 

\bibitem{Smith03}
D. R. Smith, D. Schurig, M. Rosenbluth, S. Schultz, S. Ramakrishna, and J. B. Pendry, ,,Limitations on subdiffraction imaging with a negative refractive index slab'', Appl. Phys. Lett. \textbf{82}, 10, 1506-1508 (2003).

\bibitem{Taflove}
A. Taflove and S. Hagnes, ,,Computational Electrodynamics: The Finite-Difference Time-Domain Method'', 2nd ed. (Artech House, 2000).






\end{thebibliography}
\end{document}